# Harnessing bifurcations in tapping-mode atomic force microscopy to calibrate time-varying tip-sample force measurements

Ozgur Sahin[a]

The Rowland Institute at Harvard, Harvard University, Cambridge, MA 02142

Torsional harmonic cantilevers allow measurement of time varying tip-sample forces in tapping-mode atomic force microscopy. Accuracy of these force measurements is important for quantitative nanomechanical measurements. Here we demonstrate a method to convert the torsional deflection signals into a calibrated force waveform with the use of non-linear dynamical response of the tapping cantilever. Specifically the transitions between steady oscillation regimes are used to calibrate the torsional deflection signals.

---

[a] Corresponding author: sahin@rowland.harvard.edu

# I INTRODUCTION

Tapping-mode atomic force microscopy (AFM) has enabled practical imaging of materials with nanoscale lateral resolution[1,2]. In this operation mode the force sensing cantilever vibrates at or near its resonance frequency with a large enough amplitude to avoid sticking of the tip to the sample and minimize lateral interaction. During the oscillations, attractive and repulsive forces act on the atomically sharp tip. These forces depend on the material characteristics of the sample and they affect the oscillations of the tapping cantilever. As a result, the cantilever exhibits rich non-linear dynamical behavior[3,4,5,6,7]. Several groups have investigated the cantilever dynamics in the tapping-mode in order to extract information on the material properties. While most of the initial work was on the use of vibration amplitude and phase[8,9,10,11,12], various authors are now exploring the use of high frequency vibration harmonics[13,14,15,16,17,18] and higher order resonances on the cantilever[19,20,21].

Recently, torsional harmonic cantilevers are introduced to directly measure time-varying tip-sample forces in dynamic force microscopy[22]. These cantilevers have tips that are offset from their longitudinal axis, so that tip-sample forces excite torsional vibrations (See Fig. 1). The sensitivity and high bandwidth of the torsional vibrations enable measurements of the attractive and repulsive forces and their variation with time or tip-sample separation in the tapping mode. The accuracy of these measurements is crucial for quantitative nanomechanical measurements. Therefore a proper calibration of the torsional response of the torsional harmonic cantilevers is necessary. Here we propose a method to calibrate the torsional deflection signals of torsional harmonic cantilevers. Specifically, we want to convert the electrical signals at the position sensitive detector to the tip-sample forces.

## II THEORETICAL CONSIDERATIONS

In a tapping-mode experiment performed with a torsional harmonic cantilever, raw torsional deflection signal waveform is distorted by the torsional resonance[22]. This effect is most easily understood and corrected in the frequency domain where the periodic torsional deflection signals and tip-sample forces are represented with harmonics and the response of the cantilever is represented with a frequency response function.

Harmonics of the tip-sample forces come at integer multiples of the driving frequency. Each harmonic force component results in a harmonic vibration on the cantilever in proportion to the frequency response of the torsional deflections. Considering the higher stiffness of higher order torsional modes, we approximate the torsional frequency response with a simple harmonic oscillator with a resonance frequency and quality factor equal to those of the fundamental torsional mode. With this assumption, the transfer function relating the detector signal in volts $V_T(\omega)$ to the tip-sample forces $F_{ts}(\omega)$ can be written as follows

$$V_T(\omega) = c_{optical} \frac{d}{k_T} \frac{\omega_T^2}{\omega_T^2 - \omega^2 + i\omega\omega_T/Q_T} F_{TS}(\omega).$$

(1)

Here the torsional resonance frequency and quality factor are denoted as $\omega_T$ and $Q_T$. These two parameters are easily and accurately measured in an AFM system. Torsion constant of the first torsional mode $k_T$ is defined as the angular deflection for a unit torque around the long axis of the lever. Torque is generated by the tip-sample forces acting at an offset distance $d$ from the longitudinal axis of the lever. $c_{optical}$ is the detector signal for a unit torsional deflection angle. We are neglecting the frequency response of the detector because the cut-off frequency is well above the harmonic frequencies of interest in our AFM system. $V_T(\omega)$ is the Fourier transform of the detector signal $v_T(t)$. Eq. (1) can be used to solve for $F_{TS}(\omega)$. This calculation requires the constants used in Eq. (1) to be measured. An intermediate parameter that is useful in this calculation and subsequent discussions is the corrected voltage waveform $V_{TC}$, which is defined as

$$V_{TC}(\omega) = \frac{\omega_T^2 - \omega^2 + i\omega\omega_T/Q_T}{\omega_T^2} V_T(\omega) = c_{optical} \frac{d}{k_T} F_{TS}(\omega). \tag{2}$$

Corrected voltage waveform $V_{TC}(\omega)$ can be directly calculated from the measured detector voltages with the knowledge of $\omega_T$ and $Q_T$. In time domain $V_{TC}$ has the same waveform, within a scalar factor, as $F_{TS}$. However, it remains in the electrical units (volts). Evaluation of this scalar factor is sufficient to calibrate the torsional response of torsional harmonic cantilevers.

The method of calibration presented in this letter is enabled by the fact that both torsional and vertical mechanical detection channels respond to the same time-average (DC) forces in proportion to their effective spring constants. Once the vertical spring constant is calibrated with established techniques and the time-average force $F_{ts}(\omega=0)$ is quantitatively measured by the vertical deflections, the corresponding average torsional deflection signal $V_{TC}(\omega=0)$ will be used to determine the scalar factor in Eq. (2).

In practice, DC deflection measurements of a cantilever are affected by mechanical and thermal drifts in both the cantilever and detector positions. This fact complicates the use of DC measurements in both vertical and lateral channels. However, non-linear dynamical response of the cantilever vibrations in tapping-mode provides an opportunity to eliminate drift and perform reliable calibration as we explain next.

The tapping cantilever exhibits multiple steady oscillation states depending on the drive amplitude and frequency[23]. During the transitions between different steady oscillation regimes (such as attractive and repulsive regimes), time-average forces quickly jump from a negative (attractive) value to a positive (repulsive) value. The switching time is limited by the resonance frequency and quality factor of the cantilever and it is in the millisecond time scale. Therefore, the difference between the DC deflection signals before and after a transition will be a drift free quantity. Instead of comparing the absolute values of the DC signals in the vertical and lateral channels, we are going to compare the changes in the DC values before and after an oscillatory state transition.

## III EXPERIMENTAL RESULTS

To demonstrate this calibration scheme we worked with three torsional harmonic cantilevers. Their vertical spring constants, vertical and torsional resonance frequencies, nominal width, length, and tip offset distances of these cantilevers are given in Table 1. Various methods of calibrating the vertical spring constants are reported in the literature[24,25,26]. We have used the method that calibrates the cantilever spring constant against thermo-mechanical noise, which is provided by the software running the AFM. We have obtained amplitude vs. distance and phase vs. distance curves with each cantilever on graphite. The cantilevers are driven slightly above their resonance frequencies where the free amplitude drops to approximately 60% of the resonant value. This condition favors the existence of two oscillation states[23]. Multiple cycles of amplitude vs. distance and phase vs. distance curves are plotted in Fig. 1(a-b) with results from each cantilever on a different column. At each data point DC deflections of the vertical channel is recorded and converted into force units (Fig. 1(c)). Note that there is an offset in the force values due to detector misalignment. The corresponding torsional deflection signals at each point are recorded and corrected voltage waveform $V_{TC}$ is obtained. Time average values of $V_{TC}$ are plotted in Fig 1(d).

Jumps in the phase-distance curves show the points where cantilever oscillations switch from one steady state to the other. Note that both the time-average forces in 1c and the corrected voltage waveforms in 1d exhibit jumps at the corresponding data points. A computer program is used to analyze the data and locate the jumps and record their magnitudes in Newtons for 1c and in Volts for 1d during multiple ramp cycles. The ratio of the jump magnitudes in 1c and 1d correspond to the inverse of the scalar factor in the second part of Eq. (2).

Table II gives the measured scaling factors obtained from the data in Fig. 1. After calculating $V_{TC}$ from the detector signal and taking its inverse Fourier transform, multiplication with the scaling factor will give the calibrated time-varying tip-sample force waveform.

## IV DISCUSSION

Several calibration methods for the torsional deflection signals of AFM cantilevers have been proposed and used for quantifying lateral force microscopy measurements[27,28,29,30,31,32]. Those methods that involve lateral tip-sample interactions in contact mode cannot be used directly for the calibration of torsional harmonic cantilevers, because vertical forces are also generating torque due to the offset tip. More importantly, the primary quantity that is being measured by the torsional harmonic cantilevers is the vertical tip sample force, not the lateral force. On the other hand, the torsion constant $k_T$ can be calibrated against the thermal noise. Then, according to Eq. (2), the scaling factor can be obtained by estimating $c_{optical}$ and $d$. Following this approach we calibrated $k_T$ of each cantilever against thermal noise, used $c_{optical}$ value calculated from the vertical mode force curves (we assume photo-detector gains are the same for both lateral and vertical channels), and use $d$ values in table I to estimate the scaling factor in Eq. (2). The resulting scaling factor is given in table II.

The values of the scaling factor directly calculated with the use of oscillatory state jumps and thermal noise based estimate differ as much as 50%. We believe the calibration method presented in this letter is more accurate compared to a calibration against thermal noise. First of all, it does not require the measurements of intermediate variables $k_T$, $c_{optical}$, and $d$. $c_{optical}$ and $d$ contain uncertainties ($c_{optical}$ depends on the laser spot position and $d$ is subject to misalignments in the manufacturing process). Second, thermal noise based measurement of $k_T$ values are less accurate for stiff torsional modes of tapping-mode cantilevers, because the RMS deviation in cantilever position is only slightly above the detector noise. In our method, the force jumps used for calibration are typically around 5 nN, which leads to detectible changes in the torsional deflections. The uncertainty of the calibration method presented here mainly comes from the measurement of the vertical spring constant of the cantilever. This value can be obtained more accurately (typically within 10%) with the existing calibration methods[24,25,26].

In summary we have presented a method to calibrate time-varying tip sample forces measured my torsional harmonic cantilevers operated in the tapping-mode. The method uses the fact that both vertical and lateral deflections of the torsional harmonic cantilevers respond to the same vertical tip-sample force. After calibrating the vertical deflection signals with established techniques we calibrate lateral deflection signals by comparing the time-average (DC) deflection signals in both channels. We eliminate drift related errors by using differential DC measurements near the transitions between steady oscillation states of the tapping cantilever.

**Table I. Spring constant $K_1$, resonance frequency $f_0$, torsional resonance frequency $f_T$, length $L$, width $w$, and tip offset distance $d$ of three torsional harmonic cantilevers**

|   | $K_1$ (N/m) | $f_0$ (KHz) | $f_T$ (kHz) | $L$ (μm) | $w$ (μm) | $d$ (μm) |
|---|---|---|---|---|---|---|
| 1 | 2.26 | 47.3 | 809.0 | 300 | 30 | 15 |
| 2 | 6.18 | 74.7 | 1127.0 | 275 | 30 | 25 |
| 3 | 7.97 | 59.1 | 1115.7 | 380 | 30 | 22 |

**Table II. Scaling factors for the calibration of the three torsional harmonic cantilevers. Standard deviation of the measurements are less than 5% of average values. Scaling factors calibrated against thermomechanical noise is given on the right column.**

|   | Scaling factor (nN/mV) | Scaling factor (thermomechanical) (nN/mV) |
|---|---|---|
| 1 | 2.84 | 3.38 |
| 2 | 3.61 | 5.64 |
| 3 | 10.8 | 11.8 |

**FIG. 1.** Schematic (left) and SEM picture (bottom right) of a torsional harmonic cantilever. Torsional vibrations excited due to tapping allow to measure time varying tip-sample forces (top right, curve measured on graphite).

**FIG. 2** (Color online) Amplitude vs. distance (a), Phase vs. distance (b), Average vertical force vs. distance (c), and average lateral deflection signal vs. distance (d) curves in a tapping-mode experiment performed with three cantilevers. The horizontal axis is arbitrarily referenced.

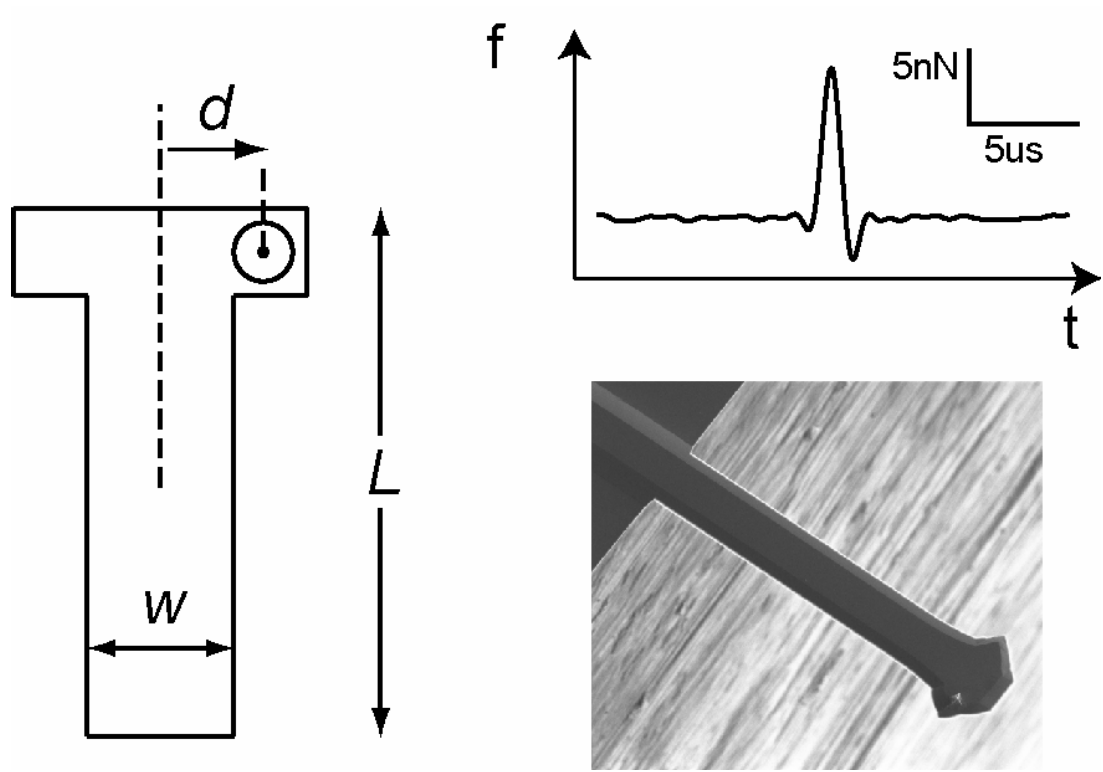

Fig.1 O. Sahin

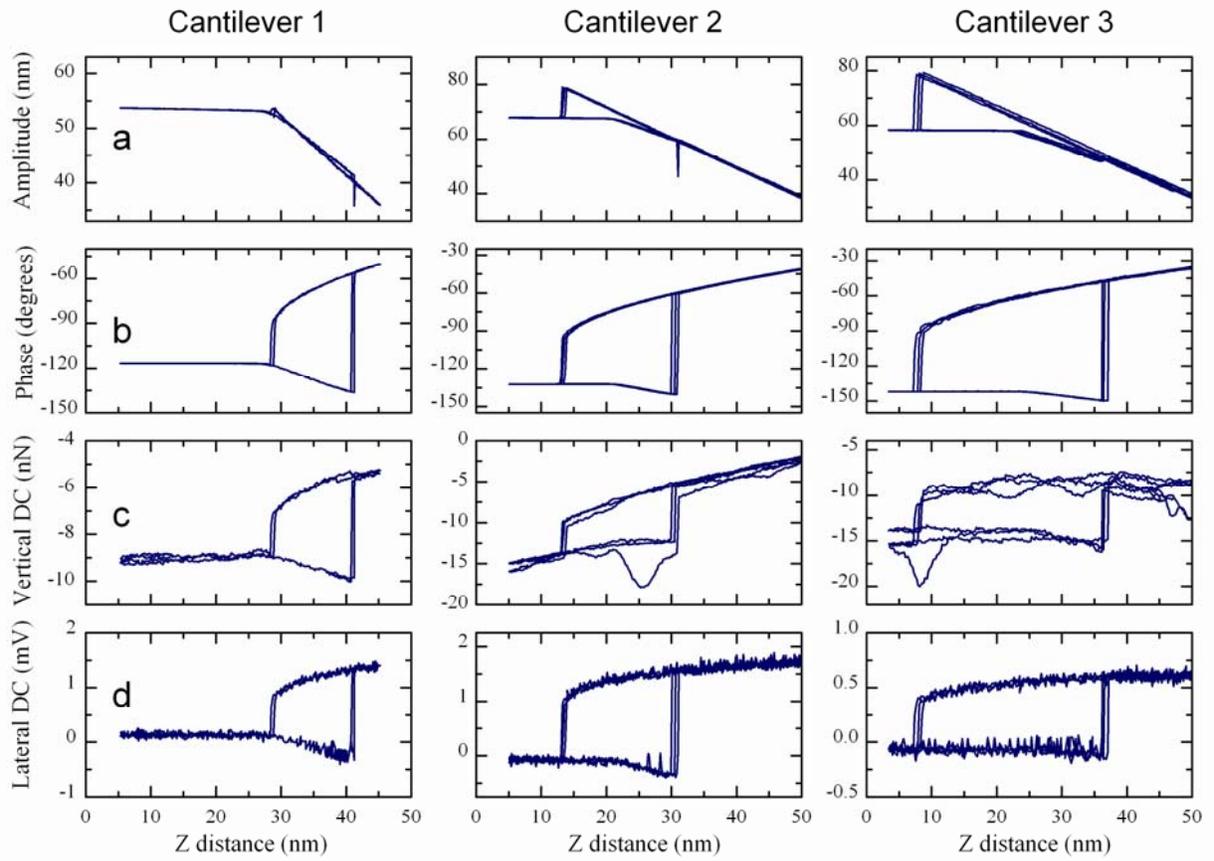

Fig. 2 O. Sahin